\documentclass{PoS}

\usepackage{graphicx}
\usepackage{bm}
\usepackage{multirow}
\usepackage{graphicx}
\usepackage{epsfig}
\usepackage{psfrag}
\usepackage{soul}
\usepackage{color}
\usepackage{multirow}
\usepackage{mathptmx}
\usepackage{mathrsfs}
\usepackage{amsmath, amssymb}
\usepackage{slashed}

\newcommand{\beq}{\begin{equation}}
\newcommand{\eeq}{\end{equation}}
\newcommand{\beqa}{\begin{eqnarray}}
\newcommand{\eeqa}{\end{eqnarray}}

\newcommand{\FslashA}[1]{\!\not{\hbox{\kern-2pt ${#1}$}}}
\newcommand{\FslashB}[1]{\!\not{\hbox{\kern+1pt ${#1}$}}}

\allowdisplaybreaks

\usepackage{dcolumn}

\title{Hierarchically deflated conjugate residual}

\ShortTitle{HDCR}

\author{\speaker{Azusa Yamaguchi}\thanks{Funded by Intel Parallel Computing Centre.}\\
  University of Edinburgh \\
  E-mail: \email{ayamaguc@staffmail.ed.ac.uk}}

\author{Peter A Boyle\\
  University of Edinburgh\\
        E-mail: \email{paboyle@ph.ed.ac.uk}}

\abstract{
We present a progress report on a 
new class of multigrid solver algorithm suitable for the solution of 5d chiral fermions such as Domain Wall fermions and the Continued Fraction overlap. Unlike HDCG \cite{Boyle:2014rwa}, the algorithm works directly on a nearest neighbour fine operator. The fine operator used is Hermitian indefinite, for example $\Gamma_5 D_{dwf}$, and convergence is achieved with an indefinite matrix solver such as outer iteration based on conjugate residual. As a result coarse space representations of the operator remain nearest neighbour, giving an 8 point stencil rather than the 81 point stencil used in HDCG. It is hoped this may make it viable to recalculate the matrix elements of the little Dirac operator in an HMC evolution.}

\FullConference{34th annual International Symposium on Lattice Field Theory\\
		24-30 July 2016\\
		University of Southampton, UK}

\begin{document}

\section{Introduction}

Despite the development of revolutionary new multilevel solver algorithms for Wilson Fermions
\cite{Luscher:2007se,Brannick:2007ue,Brannick:2007cc,Clark:2008nh,Babich:2009pc}
lying nearly ten years in the past, the extension of the approaches to all fermion
actions remains somewhat piecemeal.
The generalisation to improved Wilson (clover) fermions was made rather rapidly\cite{Osborn:2010mb},
and subsequent variations \cite{Frommer:2012mv,Frommer:2013fsa,Frommer:2013kla} have included
more efficient subspace setup.

The extension domain wall fermions\cite{Kaplan:1992bt,Shamir:1993zy} has been studied\cite{Cohen:2012sh} and an approach
made to give a substantial acceleration for valence analysis 
based on the red-black preconditioned squared operator\cite{Boyle:2014rwa}.
The stencil for the squared operator contains all points with taxicab norm less than four, giving
321 points. This has the result
that approach is unnattractive for gauge evolution code where, even if the subspace quality can be
preserved along an HMC trajectory, the reevaluation of the matrix elements of the little Dirac operator
on each timestep in the integrator, for O(50) vectors in the subspace requires naively 15000 matrix multiplies.

Even admitting a constraint, such as a minimum block size of $4^4$, the squared operator stencil only reduces
to 81 points\cite{Boyle:2014rwa}. It is clear that in order to make a practical algorithm for accelerating
HMC evolution with domain wall Fermions we must escape the constraint that the algorithm work on the squared
operator, and in order to do this we must first understand why to date only solvers making use of the squared
operator have been successful for domain wall Fermions.

\section{Spectrum of domain wall fermions}

\begin{figure}[hbt]
\includegraphics*[width=0.5\textwidth]{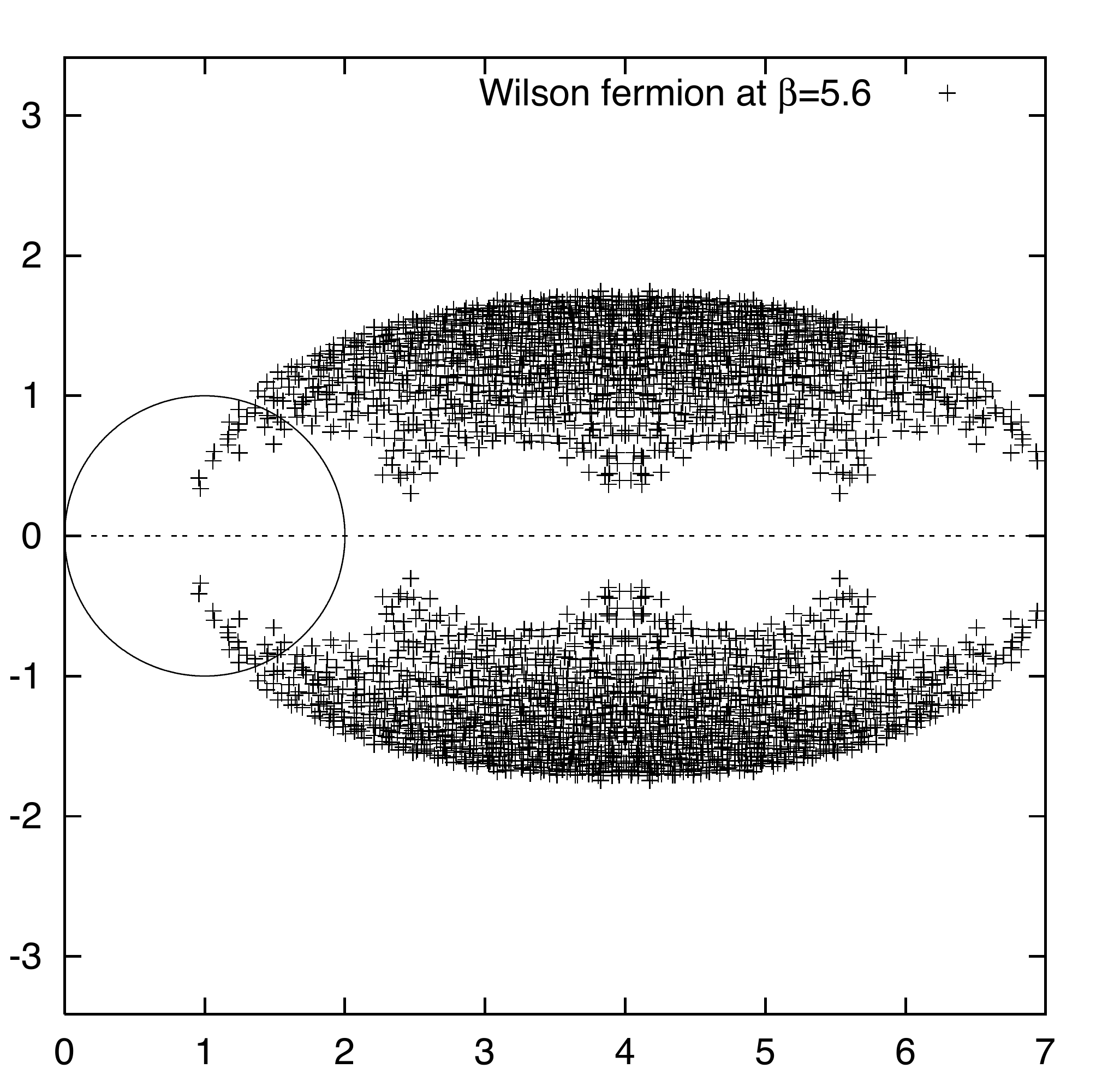}
\includegraphics*[width=0.49\textwidth]{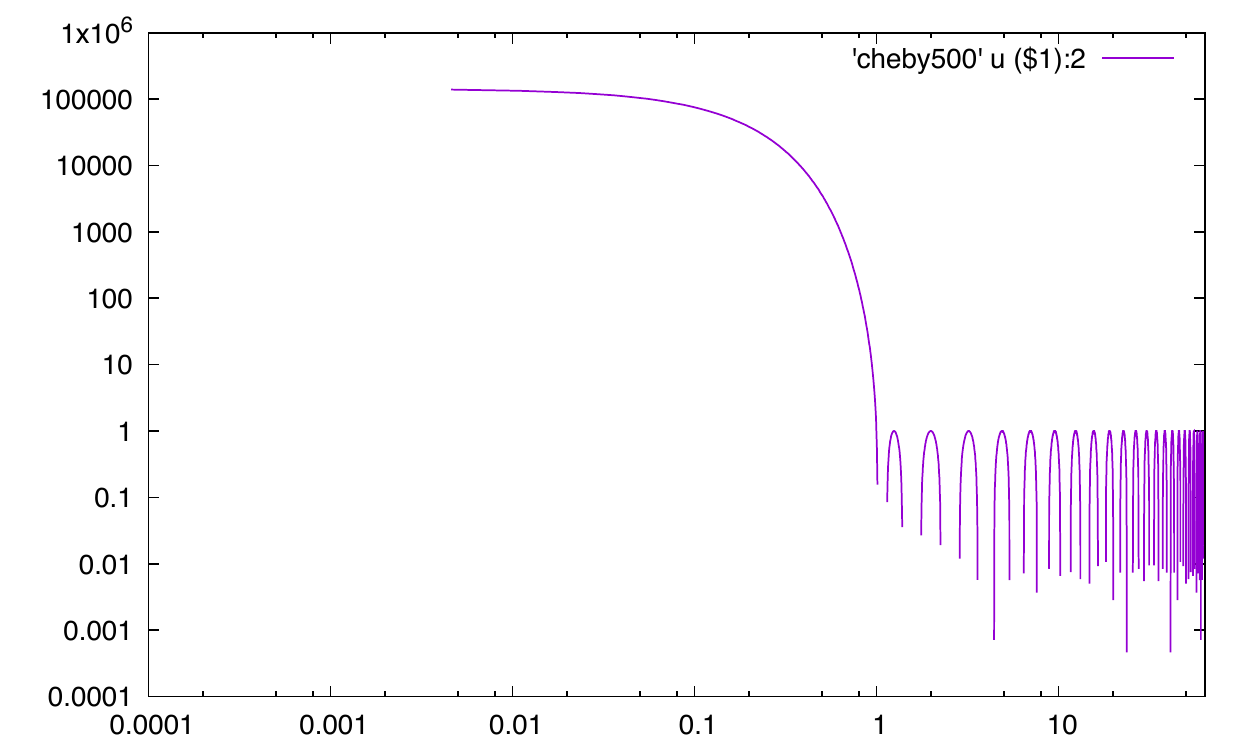}
\caption{\label{fig:hamburger}{\bf Left:} The complex eigenvalue spectrum of the Wilson operator on a $4^4$ quenched configuration with  $\beta=5.6$. This figure was produced in reference \cite{Bietenholz:2000iy}
and we simply reuse the figure here for convenience to illustrate the nature of the spectrum.
With a negative Wilson mass of order 1.5, as is introduced in the kernel of chiral fermions, the spectrum is shifted and encircles zero.
In five dimensional domain wall and related approaches a five dimensional Wilson term is introduced, without gauge links in the fifth direction, and the ``hamburger'' picture
is repeated with spectrum contained in $0 \le \Re \lambda  \le 10$ and five ``circles''.
{\bf Right:} an example of an O(500) Chebyshev low-pass filter, shifting the range of the Chebyshev to give exponentially small oscillations around zero between $[ \lambda_{\rm cut}, \lambda_{\rm max}]$,
with a five order of magnitude enhancement of low modes with $\lambda \le \lambda_{\rm cut}$.
}
\end{figure}

The spectrum of the 5d domain wall fermion operator is illustrated in figure~\ref{fig:hamburger}.
The spectrum for an appropriate negative 5d mass completely encircles and violates the \emph{folklore} present in
numerical analysis called the {\emph{half-plane condition}}\cite{trefethen}. There is a fundamental reason for this
folklore: in the infinite volume the spectrum will become dense, and the Krylov solver is then being asked to form
an (analytic) polynomial approximation to $\frac{1}{z}$ over an open region encircling the pole. It is impossible
to reproduce the phase winding around zero with an analytic function and indeed one can show that minimising
the mean square error of a fixed radius circle gives zero for all polynomial coefficients.

In the case of Conjugate Gradient on the Normal Equations (CGNE), which is used to date in RBC-UKQCD domain wall Fermion evolution,
the multiplication of each eigenvalue by its conjugate in solving
$$M_{pc}^\dag M_{pc} \psi =  \eta $$
places the phase behaviour under control and reduces the problem to a real spectrum, albeit with a squared range of eigenvalue magnitudes.

In the discrete spectrum, finite volume case, we can consider a toy models which also illustrate the problem. If the spectrum consists
of $N$ eigenvalues $\lambda_k = e^{i2\pi k/N}$ the conjugate gradient will only converge with an N-term polynomial, which can be analytically arrived
at by Gaussian elimination for small $N$.

In this paper, we propose to solve the phase problem using $\gamma_5$ Hermiticity, without squaring the operator,
leaving the coarse space representation of the operator still nearest neighbour. Since the sparsity pattern is preserved
this will represent the first true multigrid for five dimensional chiral fermions.

\section{Application to domain wall fermions}

We consider two classes of approach for chiral fermions following the nomenclature of ref. \cite{Kennedy:2006ax}. In the Cayley form,
the Hermitian indefinite operator for domain wall fermions (and Mobius fermions with $c=0$, $b\ne 1$) is 
$$H_{dwf} = \gamma_5 R_5 D_{dwf} = \Gamma_5 D_{dwf}$$
Meanwhile, the continued fraction form for the standard overlap $H_W$ kernel is already Hermitian indefinite, taking a form
that is also appropriate:
$$
\left[
\begin{array}{ccccc}
H & \frac{1}{\sqrt{\beta_0\beta_1}} & 0 & 0 &0 \\
\frac{1}{\sqrt{\beta_1\beta_0}}   & -H  & \frac{1}{\sqrt{\beta_1\beta_2}}   & 0 & 0  \\
0   & \frac{1}{\sqrt{\beta_2\beta_1}}   & H   & \frac{1}{\sqrt{\beta_2\beta_3}} & 0  \\
0   & 0   & \frac{1}{\sqrt{\beta_3\beta_2}}   & -H & \frac{1}{\sqrt{\beta_3}} \\
0   & 0   & 0   & \frac{1}{\sqrt{\beta_3}} & R\gamma_5 + \beta_0 H
\end{array}
\right].
$$
These operators are \emph{nearest neighbour} and preserve sparsity in a coarse space, but give rise to a
Hermitian \emph{indefinite} spectrum. In the infinite volume the 
spectrum will be dense, real and symmetrical about the origin. From the perspective of a Krylov solver the
polynomial approximation $P(\lambda)\sim \frac{1}{\lambda}$ must be made over a the subset
real line $\lambda\in [-\lambda_{\rm max},-\lambda_{\rm min}] \cup [\lambda_{\rm min},\lambda_{\rm max}] $

Such a spectrum succumbs easily to the conjugate residual algorithm, which relaxes the Hermitian positive definite
constraint of conjugate gradients to only Hermitian indefinite.
We will use variants of conjugate residuals as the basis of the  outer fine matrix iteration. 
Regarding the relative efficiency, it is worth to note that we create a Krylov space 
that strictly contains the CGNE Krylov space (spanned by every second term).
$$                                                                                                                                                                
P_N( D^\dag D ) D^\dag = P_N( \Gamma_5 D \Gamma_5 D ) \gamma_5 D \gamma_5                                                                                         
\subset P_{2N+1}( H_{dwf} ) \gamma_5 =  P_{2N+1}( \Gamma_5 D ) \gamma_5                                                                                           
$$
Further, since either on average or in the infinite volume, the spectrum will be symmetrical about zero,
the even terms cannot contribute to an approximation of the (odd) function $\frac{1}{x}$ and the in this limit the 
iteration should converge with an identical number of applications of the 
nearest neighbour fermion operator as unpreconditioned CGNE. This rule is observed to be almost exactly true even on $16^3$
configurations.

\section{Two level preconditioner}

To introduce a Krylov process as a multigrid preconditioner, we use variable preconditioned GCR as the outer iteration.
Since this is a stadard algorithm we do not document it in the interests of brevity.
Multigrid may now introduced as the \emph{Preconditioner}. We have
tried several approaches to define the low mode vectors used in coarsening. These included
i) inverse iteration applied to Gaussian noise, ii) Lanczos eigenvectors, and iii) Chebyshev filters applied to Gaussian noise.
An example of our use of high-order Chebyshev filters is given in figure~\ref{fig:hamburger}. The rapid divergence of a high order
Chebyshev outside the default interval $[-1,1]$ is used to \emph{enhance} the modes of interest. We adopt the trick from polynomial
preconditioned implicitly restarted Lanczos \cite{rudy}.
Having obtained a basis that captures the near null space of the operator, the vectors are projected into left handed and right handed
chiralities. This $\gamma_5$ compatible approach was important to eliminate near zero eigenvalues in the coarsened operator\footnote{Suggested
to the authors by Kate Clark.}.

The vectors $\phi_k$ are then restricted to blocks, of size $2^4$ in space time, and the full extent of the fifth dimension, enabling a
coarse space representation to be built up as follows.
\beq
\phi^b_k(x) = \left\{ \begin{array}{ccc}
  \phi_k(x) &;& x\in b\\
  0 &;& x \not\in b
\end{array}
\right.
\eeq
\beq{\rm span} \{ \phi_k\}\subset
{\rm span} \{ \phi_k^b\} .\eeq
\beq
P_S =  \sum_{k,b} |\phi^b_k\rangle \langle \phi^b_k | \quad\quad ; \quad\quad P_{\bar{S}} = 1 - P_S
\eeq
\beq
M=
\left(
\begin{array}{cc}
M_{\bar{S}\bar{S}} & M_{S\bar{S}}\\
M_{\bar{S}S} &M_{SS}
\end{array}
\right)=
\left(
\begin{array}{cc}
P_{\bar{S}} M P_{\bar{S}}  &  P_S M P_{\bar{S}}\\
 P_{\bar{S}} M P_S &   P_S M P_S
\end{array}
\right)
\eeq
we can represent the matrix $M$ exactly on this subspace by computing its matrix elements,
known as the \emph{little Dirac operator} (coarse grid matrix in multi-grid)
\beq
A^{ab}_{jk} = \langle \phi^a_j| M | \phi^b_k\rangle
\quad\quad ; \quad\quad
(M_{SS}) = A_{ij}^{ab} |\phi_i^a\rangle \langle \phi_j^b |.
\eeq
the subspace inverse can be solved by Krylov methods and is:
\beq
Q =
\left( \begin{array}{cc}
0 & 0 \\ 0 & M_{SS}^{-1}
\end{array} \right)
\quad\quad ; \quad\quad
M_{SS}^{-1} = (A^{-1})^{ab}_{ij} |\phi^a_i\rangle \langle \phi^b_j |
\eeq

It is important to note that $A$ inherits a sparse structure from $M$ because well separated blocks do \emph{not} connect through $M$.
We can Schur decompose the matrix
\begin{eqnarray*}
M= U D L = \left[ \begin{array}{cc}M_{\bar{s}\bar{s}} & M_{\bar{s}s} \\ M_{s\bar{s}} & M_{ss} \end{array} \right]
&=&
\left[ \begin{array}{cc} 1 & M_{\bar{s} s}  M_{ss}^{-1} \\ 0 & 1 \end{array} \right]
\left[ \begin{array}{cc} S & 0 \\ 0 & M_{ss} \end{array} \right]
\left[ \begin{array}{cc} 1 & 0 \\ M_{ss}^{-1} M_{s \bar{s}} & 1 \end{array} \right]
\end{eqnarray*}
Note that
$P_L M = \left[ \begin{array}{cc} S & 0 \\ 0 &0 \end{array} \right]$ yields the Schur complement $ S = M_{\bar{s}\bar{s}} - M_{\bar{s}s} M^{-1}_{ss} M_{s\bar{s}}\
 $,
and that the diagonalisation $L$ and $U$ are projectors $P_L$ and $P_R$ (Galerkin oblique projectors in multi-grid)
\beq
P_L = P_{\bar S} U^{-1} =\left( \begin{array}{cc}
 1 & -M_{\bar{S} S}  M_{SS}^{-1}\\
 0 & 0
             \end{array} \right)
\quad\quad ; \quad\quad
P_R = L^{-1} P_{\bar{S}}  =
\left( \begin{array}{cc}
1 & 0 \\ -M_{SS}^{-1} M_{S \bar{S}} & 0
\end{array} \right)
\eeq
We introduce a smoother which is an order 10 Chebyshev polynomial approximation to $1/x$ in the range $[1.0,64.0]$.
To maintain hermiticity in the outer iteration, we presently introduce the smoother and coarse grid preconditioner in a symmetric way, with the composite
outer Krylov operating on the matrix as documented in \cite{Boyle:2014rwa}:
$$M_{\rm outer} = M_{chebyshev} P_L + P_R M_{chebyshev} + Q - M_{chebyshev} P_L {\cal H}  M_{chebyshev}.$$

\section{Initial results}

We use a standard RBC-UKQCD $a^{-1} = 1.73$ GeV ensemble with the Iwaski gauge action and DWF 2+1 dynamical flavours with
light mass $a m_l = 0.01$ and strange mass $a m_s = 0.03$ and volume $16^3\times 32 \times 16$. 
To make a viable test system, we set the valence mass artificially low to 0.001 to increase the condition
number, resulting in thousands of conjugate gradient iterations. 
We use 16 nodes on Cori phase-1 at NERSC, and take 16 subspace vectors and an order O(900) polynomial.

We display present results from the $16^3$ configuration in table~\ref{tab:timing}.
A speed up of around a factor
of three is obtain in the solution time even form the small volume system. The set up time still presently
exceeds the original solve time. The relative speed up is expected to grow as our
study progresses to even less well conditioned systems but is the subject of further study.

While it is certainly not yet clear that the final algorithm will be applicable for use in Hybrid Monte Carlo,
there are reasons for encouragement. 
The Lanczos vectors and Chebyshev filtered vectors both demonstrate real speed up over the original
red-black conjugate gradient, despite not yet deflating the coarse grid operator. One or two stages of inverse
iteration did not yield competitive solution times and appeared less promising as a subspace setup approach.
The 40s solve time was composed of 27s on the fine operator (smoother) and 13s on the coarse space.
The coarse space is presently consuming $\frac{1}{3}$ of the time, but
has not itself received any further deflation and the algorithm remains strictly two level. Our code
implementation in Grid is in principle recursive, and either true recursive multigrid or coarse space eigenvector
deflation are open options. Once the coarse space is made cheaper cost can be rebalanced by 
solving more exactly and using more vectors for the coarse space.

The 300s Chebyshev setup is too long on the test system to be used in HMC on this volume and mass; however
since the Chebyshev polynomial is evaluated through a recurrence relation it is also possible to generate
Chebyshevs with many different orders for fixed cost. This avenue has not yet been explored. Further,
it has become common in multigrid to use polynomial prediction or other schemes to track the subspace across an HMC trajectory
since the motion of the gauge configuration field space is limited by the step size, so it is possible this cost
could be amortised across a trajectory rather than a single solution.

\begin{table}[hbt]
\begin{tabular}{cccc}
Algorithm & setup/vecs &Fine Matmuls & Time\\
CGNE      &      -      & 3221   & 110s \\
HDCR      &  Lanczos/16 &    & 45s\\
HDCR      &  $M^{-1}$/16  &    & 120s\\
HDCR      &  Cheby/16   &    & 40s\\
\hline
Coarse    &             &    & 13s\\
Fine      &             & 624& 27s\\
\hline
Chebyshevs&             &    & 300s
\end{tabular}
\caption{ \label{tab:timing} We display the wall clock timing and matrix multiply count for HDCR. 
}
\end{table}

\section{Acknowledgements}

A.Y. has been supported by an Intel Parallel Computing Centre held at the Higgs Centre for Theoretical Physics.
P.B. acknowledges Wolfson Fellowship WM160035, an Alan Turing Fellowship, and 
STFC grants ST/P000630/1, ST/M006530/1, ST/L000458/1, ST/K005790/1, ST/K005804/1, ST/L000458/1.
We would like to thank Martin Luescher, Kate Clark, Evan Weinberg and Richard Brower for useful discussions.
All code has been implemented in the Grid library\cite{Boyle:2016lbp}, and the Cori phase-1 system at NERSC/LBNL has been
used for the code development and testing.

\end{document}